\documentclass[aps,pra,reprint,amsmath,amssymb,superscriptaddress,showpacs,nobibnotes]{revtex4-1}
\usepackage{graphicx,SIunits,mathrsfs}
\usepackage{bm}     
\usepackage{hyperref} 
\usepackage{dsfont}    
\usepackage{color}

\begin{document}

\title{Efficient generation of multipartite W state via quantum eraser}

\author{Yong-Su Kim}
\email{yong-su.kim@kist.re.kr}
\affiliation{Center for Quantum Information, Korea Institute of Science and Technology (KIST), Seoul, 02792, Republic of Korea}
\affiliation{Division of Nano \& Information Technology, KIST School, Korea University of Science and Technology, Seoul 02792, Republic of Korea}

\author{Young-Wook Cho}
\affiliation{Center for Quantum Information, Korea Institute of Science and Technology (KIST), Seoul, 02792, Republic of Korea}

\author{Hyang-Tag Lim}
\affiliation{Center for Quantum Information, Korea Institute of Science and Technology (KIST), Seoul, 02792, Republic of Korea}

\author{Sang-Wook Han}
\affiliation{Center for Quantum Information, Korea Institute of Science and Technology (KIST), Seoul, 02792, Republic of Korea}

\date{\today} 

\begin{abstract}
\noindent  

Entanglement among multiple particles is a keystone for not only fundamental research on quantum information but also various practical quantum information applications. In particular, W state has attracted a lot of attention due to the robustness against particle loss and the applications in multiparty quantum communication. However, it is challenging to generate photonic W state with large number of photons since $N$-photon W state requires superposition among $N$ probability amplitudes. In this paper, we propose an efficient linear optical scheme to generate $N$-photon W state via quantum erasure. The success probability of our protocol polynomially decreases as the number of photons increase. We also discuss the experimental feasibility of our protocol, and anticipate that one can efficiently generate tens of photonic W state with our scheme using currently available quantum photonics technologies.

\end{abstract}


\maketitle


\noindent Entanglement is at the heart of quantum information~\cite{EPR,horodecki09}. In particular, entanglement among multiple qubits plays crucial roles in quantum information processing such as nonlocality test~\cite{brunner14}, multi-party quantum communication~\cite{hillery99}, and quantum computation~\cite{ladd10}. Two representative genuine multipartite entanglement, the Greenberger-Horne-Zeilinger (GHZ) states and the W states, have significantly different features. Note that the interchange between these two states via local operation and classical communication is forbidden~\cite{dur00}.

The $N$-qubit GHZ sate is represented as 
\begin{equation}
|{\rm GHZ}_N\rangle=\frac{1}{\sqrt{2}}\left(|0\rangle^{\otimes N}+|1\rangle^{\otimes N}\right).
\end{equation}
Since the GHZ state maximally violates the Bell-type inequality, it is usually considered as a  maximally entangled state. However, it is extremely fragile, so easily looses entanglement. For instance, if one or more qubit particles are lost, the remaining $M$-qubit state becomes a statistical mixture of 
\begin{equation}
\rho_M=\frac{1}{2}\left(|0\rangle\langle0|^{\otimes M}+|1\rangle\langle1|^{\otimes M}\right).
\end{equation}
In experiment, GHZ states up to tens of qubits have been observed in photonic qubits~\cite{wang16,wang18,zhong18}, trapped ions~\cite{monz11}, and superconducting qubits~\cite{kelly15,song17}.

Another representative genuine multipartite entangled state, the W state, is represented as
\begin{equation}
|{\rm W}_N\rangle=\frac{1}{\sqrt{N}}\left(|0\cdots01\rangle+|0\cdots010\rangle+\cdots+|10\cdots0\rangle\right).
\label{W}
\end{equation}
It is notable that, in the W state, every qubit shares the optimal amount of entanglement with all other qubits~\cite{koashi00,liu02}. This unique property suggests a web-like structure that every qubit is coupled with all other qubits. Therefore, even if one or more qubits are lost, the remaining $M$-qubit state maintains the form of W state, $|{\rm W}_M\rangle$.

Photonic qubit implementation of W state provides fundamental quantum information test platforms~\cite{zhang16} as well as practical applications to multiparty quantum communication~\cite{joo04,cao06,wang07,zhu15,lipinska18}, and quantum metrology~\cite{fra16}. There has been a few theoretical proposals~\cite{zou02,yamamoto02,shi05,krenn17,blasiak18,gu19,feng19} and experimental implementations~\cite{eibl04,mikami05,tashima09,tashima10,fang19} of photonic W state generation. However, W state preparation with large number of photons is extremely challenging since the number of probability amplitudes increases as the photon number increases. Due to the difficulty, there are only few proposals to generate W state with arbitrary number of qubits from independent single photons~\cite{lim05,tashima09-b,ikuta11,hu15}. However, these schemes suffer from low success probability, i.e., the typical success probability $P_N$ to generate $N$-qubit W state exponentially decreases as $N$ increases, $P_N\sim \mathcal{O}^{-N}$.




In this paper, we propose an efficient linear optical protocol to generate $N$-photon W state from single photons via quantum eraser. We found that the success probability of our scheme approaches to $P_N\sim 1/N$ for a large $N$, and thus shows much higher than those of other conventional W state generation schemes~\cite{lim05,tashima09-b,ikuta11,hu15}. It is also remarkable that the qubit particles have never overlapped during the entangling process with each other. It supports a counter-intuitive result that the qubit particles do not need to touch one another for entanglement generation~\cite{yurke92-a,yurke92-b,kim18,blasiak18}. 



\begin{figure}[t]
\centering
\includegraphics[width=3.4in]{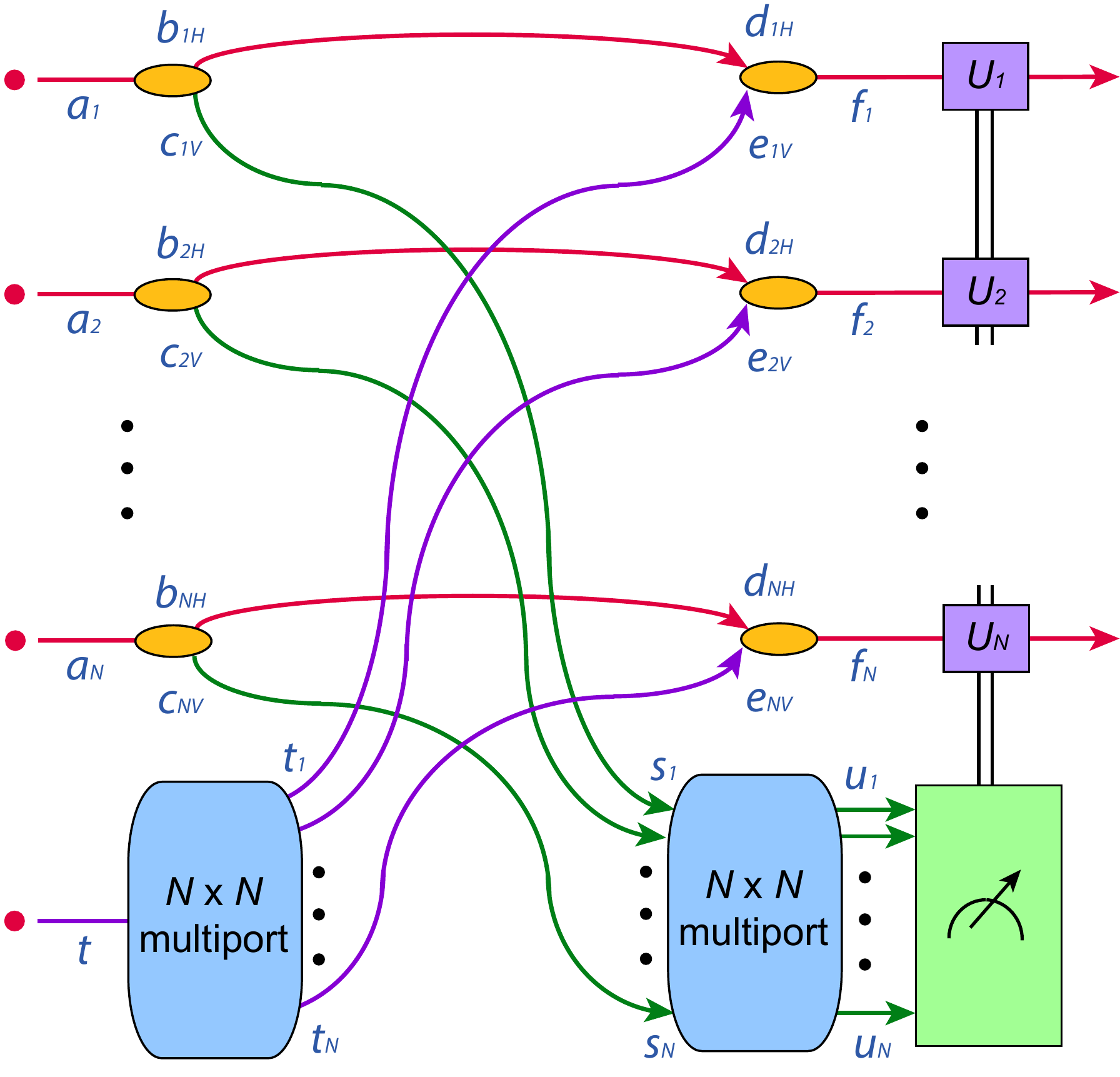}
\caption{Our scheme to generate $N$-photon W state via quantum eraser.}
\label{scheme}
\end{figure}

Figure~\ref{scheme} shows our schematic to generate $N$-qubit W state. For simplicity, we present the scheme with photonic polarization qubits, however, it is valid for any type of bosonic particles. It can be also applied for other degrees of freedom of photons such as discrete-energy entanglement which is preferred for long distance communication~\cite{fang19}. The protocol begins with $N+1$ single qubit inputs of $|\psi\rangle_{{\rm in}}=a_1^{\dag}a_2^{\dag}\cdots a_N^{\dag}\otimes t^{\dag}|0\rangle$ where $a_{i}^{\dag}$ and $t^\dag$ denote creation operators at $a_i$ and $t$, respectively. Then, the input state of $a_i$ and $t$ are divided by polarizing beamsplitters (PBS) and a symmetrical $N\times N$ multiport, respectively. The transformations can be presented as
\begin{eqnarray}
a_i^{\dag}&\rightarrow& \alpha_i b_{iH}^{\dag}+\beta_i c_{iV}^{\dag},\nonumber\\
t^{\dag}&\rightarrow& \frac{1}{\sqrt{N}}\left(t_1^{\dag}+t_2^{\dag}+\cdots+t_N^{\dag}\right),
\label{input}
\end{eqnarray}
where the subscript $H~(V)$ denotes horizontal (vertical) polarization, and $\alpha_i$ and $\beta_i$ are complex coefficients which satisfy $|\alpha_i|^2+|\beta_i|^2=1$. Here, we assume that all the relative phase between $t_i$ are zero for simplicity. Note that the coefficients $\alpha_i$ and $\beta_i$ can be controlled by changing the input polarization state of $a_i$. For simplicity, we will assume the probability 
\begin{equation}
p_h=|\alpha_i|^2
\label{p_h}
\end{equation}
for all $i$. Then, as shown in Fig.~\ref{scheme}, we take the transformation of 
\begin{eqnarray}
b_{iH}^{\dag}&\rightarrow& d_{iH}^{\dag}\rightarrow f_{iH}^{\dag},\nonumber\\
c_{iV}^{\dag}&\rightarrow& s_i^{\dag} \rightarrow \sum_j \gamma_{ij}u_j^{\dag},\\
t_i^{\dag}&\rightarrow& e_{iV}^{\dag}\rightarrow f_{iV}^{\dag}.\nonumber
\label{transform}
\end{eqnarray}
where $\gamma_{ij}$ are the complex coefficients which satisfy $\sum_j|\gamma_{ij}|^2=1$. The transformation $s_i^{\dag} \rightarrow \sum_j \gamma_{ij}u_j^{\dag}$ presents $N$-port interference by a symmetrical $N\times N$ multiport with a $s_i$ input.

For W state generation, we post-select the case when all $N+1$ outputs, i.e., all $f_i$ and one of $u_j$, are occupied by single photon states. From the transformation configuration, it happens only when one of $a_i^{\dag}$ takes $c_{iV}^{\dag}$ while all other $a_j^{\dag}$ take $b_{jH}^{\dag}$, and $t^{\dag}$ takes $t_i^{\dag}$, respectively. For example, if $a_1^{\dag}$ takes $a_1^{\dag}\rightarrow c_{1V}^{\dag}$, $a_j^{\dag}$ where $j\neq1$ should take $a_j^{\dag}\rightarrow b_{jH}^{\dag}$ and $t^{\dag}\rightarrow t_1^{\dag}$. In this case, the output state is presented as 
\begin{equation}
|\Psi_1\rangle=|\psi_1\rangle\otimes|s_1\rangle=f_{1V}^{\dag}f_{2H}^{\dag}\cdots f_{NH}^{\dag}\otimes s_1^{\dag}|0\rangle.
\label{Psi_1}
\end{equation}
Note that in this state presentation, we do not take the transformation of $s_i^{\dag} \rightarrow \Sigma_j \gamma_{ij}u_j^{\dag}$, yet. Similarly, if $j$-th input takes $a_j^{\dag}\rightarrow c_{jV}^{\dag}$, the post-selected output state is given as
\begin{eqnarray}
|\Psi_j\rangle&=&|\psi_j\rangle\otimes|s_j\rangle\nonumber\\
&=& f_{1H}^{\dag}\cdots f_{(j-1)H}^{\dag}f_{jV}^{\dag}f_{(j+1)H}^{\dag}\cdots f_{NH}^{\dag} \otimes s_j^{\dag}|0\rangle.
\label{outputs}
\end{eqnarray}
Since $|\psi_j\rangle$ are distinguishable each other due to $|s_j\rangle$, the overall $N$-qubit state $\rho_f$ is given as a mixture of $|\psi_j\rangle$ of
\begin{eqnarray}
\rho_f=\frac{1}{N}\left(|\psi_1\rangle\langle\psi_1|+\cdots+|\psi_N\rangle\langle\psi_N|\right).
\label{mixture}
\end{eqnarray}

In order to make the overall state $\rho_f$ pure, which corresponds to the coherent superposition of $|\psi_j\rangle$, we need to employ a quantum eraser to delete the spatial mode information of $|s_j\rangle$~\cite{kim00, neves09}. By employing a symmetrical $N\times N$ multiport and measuring one of the outputs $u_k$, the final state becomes a coherent superposition of $|\psi_j\rangle$ as

\begin{eqnarray}
|W_N^{(k)}\rangle=&&\frac{1}{\sqrt{N}}\sum_{j=1}^{N}e^{i\phi_j^{(k)}}|\psi_j\rangle\nonumber\\
=&&\frac{1}{\sqrt{N}}\big(e^{i\phi_1^{(k)}}f_{1V}^{\dag}f_{2H}^{\dag}\cdots f_{NH}^{\dag}\nonumber\\
&&+e^{i\phi_2^{(k)}}f_{1H}^{\dag}f_{2V}^{\dag}f_{3H}^{\dag}\cdots f_{NH}^{\dag}+ \cdots \nonumber\\
&&+e^{i\phi_N^{(k)}}f_{1H}^{\dag}\cdots f_{(N-1)H}^{\dag}f_{NV}^{\dag}\big)|0\rangle,
\label{W_state}
\end{eqnarray}
Here, $\phi_j^{(k)}$ denotes the relative phase between $|\psi_j\rangle$ when the ancillary qubit is measure at $u_k$. Note that $|W_N^{(k)}\rangle$ becomes a $N$-qubit W state as long as the relative phase $\phi_j^{(k)}$ is fixed. Since the relative phase can be fixed with the detection of the ancillary qubit at $u_k$, it heralds $N$-qubit W state preparation at the output modes $f_j$. Note that the different ancillary qubit detection at $u_k'$ implies different relative phase $\phi_j^{(k')}$. The different relative phase can be adjusted with the local unitary operations of $U_1^{(k)}\otimes\cdots\otimes U_N^{(k)}$ to the $N$ qubits. Therefore, one can increase the success probability of the W state generation scheme by performing the feedforward unitary operations $U_1^{(k)}\otimes\cdots\otimes U_N^{(k)}$ at $N$ qubits according to the ancillary qubit detection result $u_k$.
 

\begin{table}[t]
\begin{center}
\begin{tabular}{c|c}\hline
Protocols	& ~~Success probability, $P_N$~~    \\\hline
Using multiport~\cite{lim05}	& $e^{1.35-1.27N}$ \\\hline
Quantum fusion~\cite{tashima09-b}	& $N/5^{N-1}$ \\\hline
Fusion with X-phase~\cite{hu15}	& $(N+1)/2^{N}$ \\\hline
~Ours without feedforward~	& $(N-1)^{N-1}/N^{N+1}$ \\\hline
Ours with feedforward	& $(N-1)^{N-1}/N^N$ \\\hline
\end{tabular}\caption{The success probability of various protocols for generating $N$-photon W state.}
\end{center}
\label{compare}
\end{table}

\begin{figure}[b]
\centering
\includegraphics[width=3.4in]{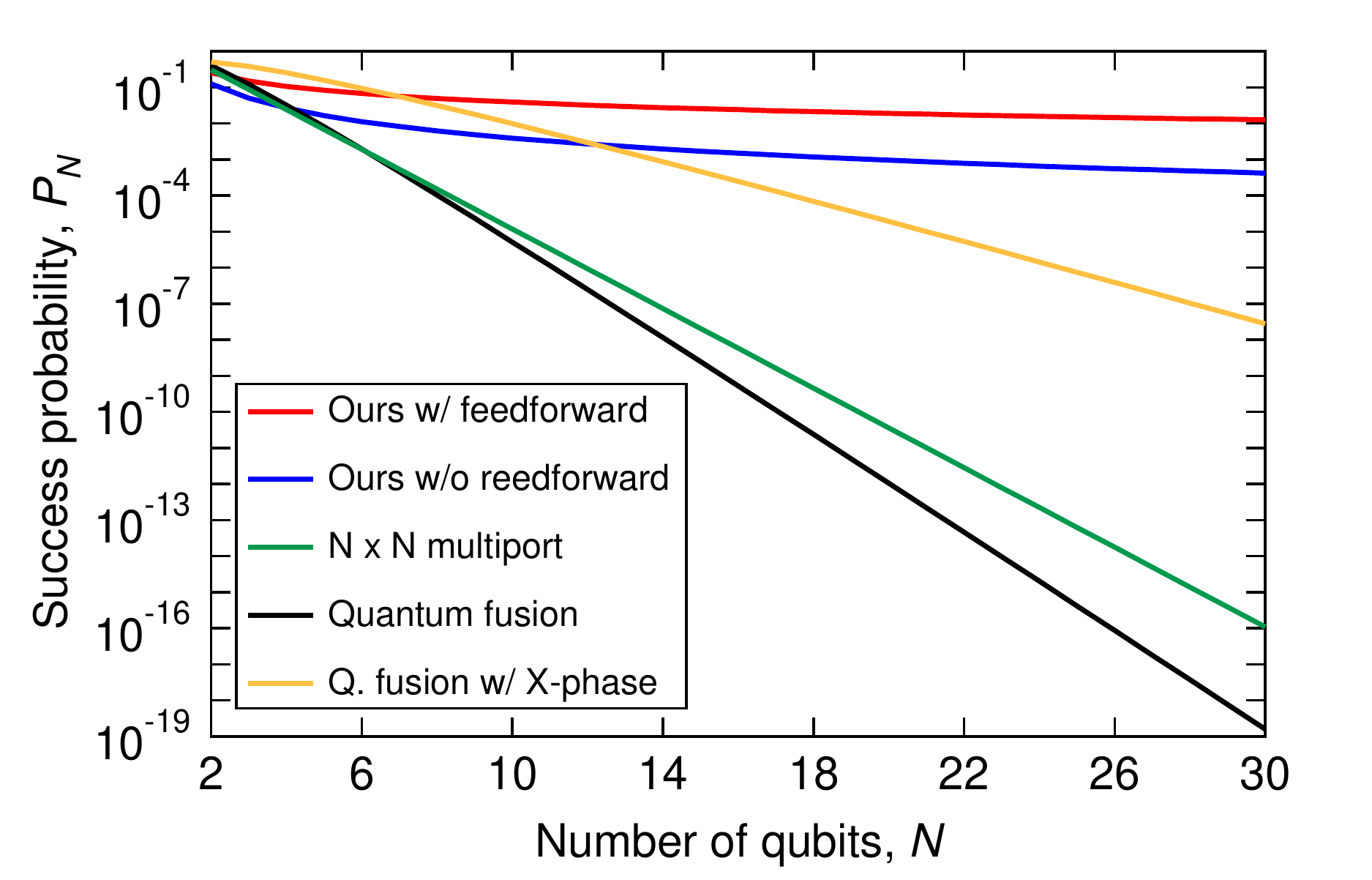}
\caption{The success probability of W state generation with respect to the number of qubits. Ours w/ (w/o) feedforward stands for our protocol with (without) feedforward. ${\rm N}\times {\rm N}$ multiport denotes the scheme using ${\rm N}\times {\rm N}$ entanglement multiport~\cite{lim05}. Quantum fusion denotes $N$-photon W state generation using quantum fusion method which generates $|W\rangle_N$ from $|W\rangle_{N-1}$~\cite{tashima09-b}. Q. fusion w/ X-phase presents quantum fusion method with cross-phase modulation which is nonlinear optical effect~\cite{hu15}.}
\label{result}
\end{figure}

Let us discuss the success probability of our W state generation scheme. The simultaneous probability of $a_i^{\dag}\rightarrow c_{iV}^{\dag}$, $a_j^{\dag}\rightarrow b_{jH}^{\dag}$ for all other $j\neq i$, and $t^{\dag}\rightarrow t_i^{\dag}$ is given as 
\begin{equation}
P=\frac{1}{N}(1-p_h)p_h^{N-1}.
\end{equation}
Considering $N$ qubits which can take $c_{iV}$ and the probability $1/N$ for measuring $u_k$, the overall success probability of obtaining W state remains the same. Therefore, with the condition of Eq.~(\ref{p_h}), the maximum success probability of generating $N$-qubit W state, \eqref{outputs} is given as
\begin{equation}
P_N^{(k)}=\max_{p_h}P=\frac{(N-1)^{N-1}}{N^{N+1}},
\label{prob2}
\end{equation}
for $p_h=\frac{N-1}{N}$. Note that for a large $N$, the maximum success probability becomes 
\begin{equation}
\lim_{N\to\infty}P_N^{(k)}\rightarrow\frac{1}{N^2}.
\label{prob3}
\end{equation}

It is remarkable that the success probability can be further increased by performing feedforward operations of $U_i^{(k)}$ according to the measurement result $u_k$. In this case, we can obtain the unity probability for measuring $u_k$ instead of $1/N$, and thus, the success probability becomes
\begin{equation}
P_{N}^{{\rm FF}}=\frac{(N-1)^{N-1}}{N^{N}}.
\label{prob_FF}
\end{equation}
For a large $N$, the maximum success probability with feedforward operation becomes
\begin{equation}
\lim_{N\to\infty}P_{N}^{{\rm FF}}\rightarrow\frac{1}{N}.
\label{prob3}
\end{equation}
Overall, the success probability of of our protocol to W state generation polynomially decreases as the number of photons $N$ increases.

We compare the success probabilities of representative protocols for generating $N$-photonic qubit W state in Table~1. Note that the success probabilities of other protocols exponentially decrease ($P_N\sim\mathcal{O}^{-N}$) as the number of photons $N$ increases which are much faster that of our protocol. The success probability as a function of number of qubits $N$ is presented in Fig.~\ref{result}. It is remarkable that the success probability of our protocol is significantly higher than other representative protocols to generate W state with tens of photons. 

\begin{figure}[b]
\centering
\includegraphics[width=3.4in]{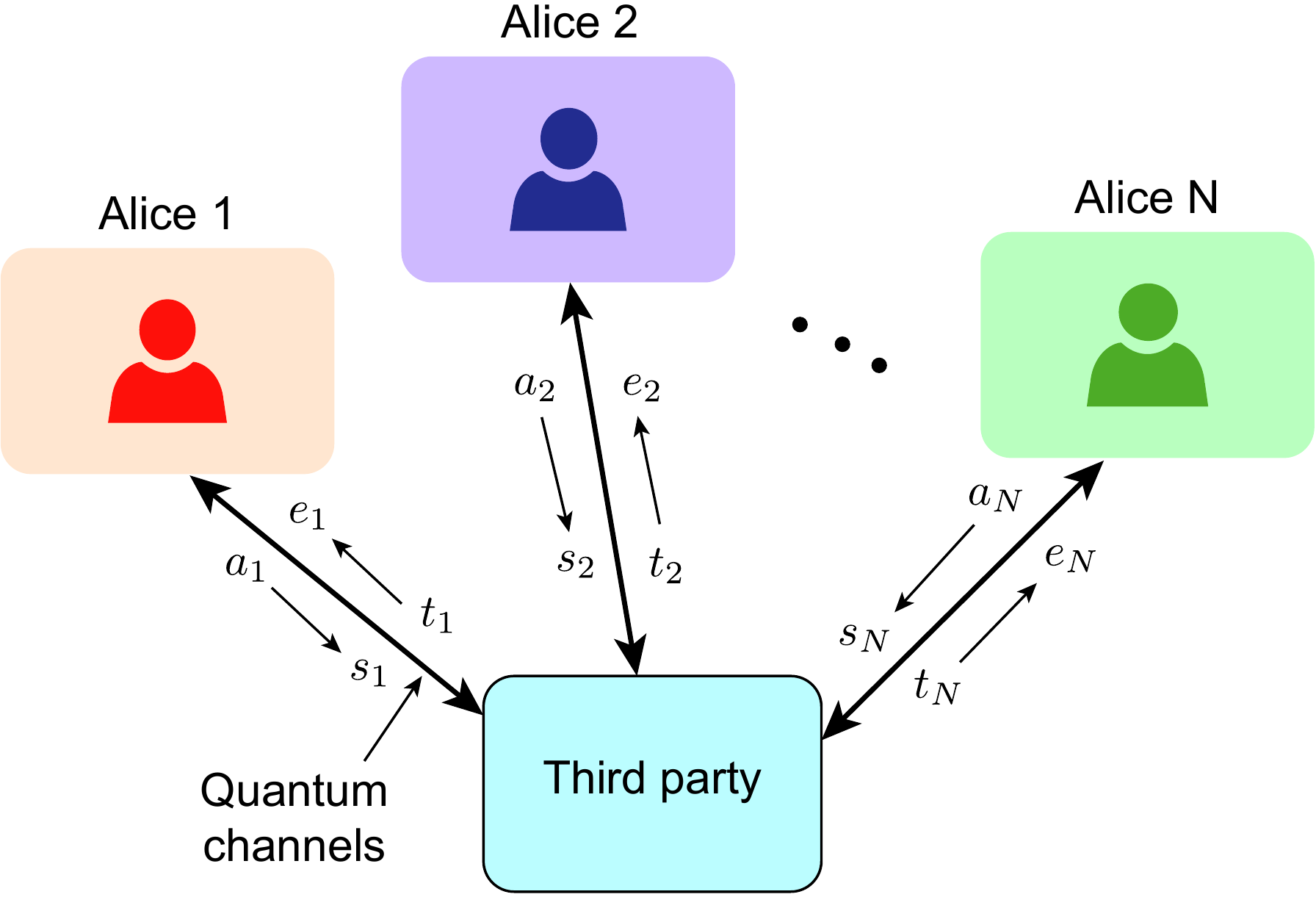}
\caption{Informationally balanced W state generation among distant multiple parties.}
\label{comm}
\end{figure}


The transformation and the post-selection of successful cases guarantee that the indistinguishable photons have never overlapped during the entangling process. The absence of photon overlap implies that the entangling process does not happen at a well-defined region. Rather it can be shared by multiple separated regions~\cite{kim18,blasiak18}. 

As an application of the non-overlapping particle property, one can implement informationally balanced W state generation among distant parties~\cite{kim18,tanu19}. Figure~\ref{comm} shows the conceptual scheme for informationally balanced W state generation among distant multiple parties. Each communication party possesses a single-photon source and detector. While the mode $b_{iH}$ is kept, they send $c_{iV}$ to the third party via quantum channels ($c_i\rightarrow s_i$). The third party, who also has a single-photon source and detector, receives photonic modes $c_i$ from the communication parties. At the same time, she transmits modes $t_i$ to the communication parties via quantum channels ($t_i\rightarrow e_i$). After the necessary unitary transformations at the communication parties and the third party, the distant multiple communication parties share W state only when all the communication parties and the third party have a single-photon each. Note that the success probability of the W state generation can be increased with the announcement of third party detection result $u_k$ and following unitary operations at the communication parties $U_k$. Unlike the GHZ state generation, it requires a third party to announce the measurement result of $u_k$, however, she does not participate in sharing qubit particles. Note that the information balance is a critical feature in most quantum communication applications~\cite{shamir, kim18, tanu19}.


In order to generate W state using our scheme, one needs $N+1$ identical single-photon inputs, and linear optical network with excellent phase stability. We remark that it is already possible to generate tens of photonic qubit W state generation with currently available quantum photonic technologies. For instance, it has been reported that generation of 12 identical single-photon states using spontaneous parametric down-conversion~\cite{zhong18}. Recently, 20 identical single-photon state generation using a quantum dot has been presented~\cite{wang19}. The complicated linear optical network can be implemented with rapidly developing integrated quantum photonics which also provides high level of phase stability~\cite{wang19-b}. Note that our protocol which is presented with photonic polarization modes can be easily converted to, for example, discrete-energy modes~\cite{fang19}, so it can be directly applied to integrated quantum photonics.


In summary, we have proposed an efficient linear optical protocol to generate multi-photon W state via quantum eraser. We have found that the success probability of our protocol polynomially decreases ($P_N\sim N^{-2}$ without feedforward and $P_N\sim N^{-1}$ with feedforward) as the number of photons $N$ increases. Considering the success probability of other representative photonic W state generation methods exponentially decreases ($P_N\sim\mathcal{O}^{-N}$) as $N$ increases, our protocol provides a powerful tool to investigate quantum information with multipartite entanglement. We remark that our protocol can be implemented with the current quantum photonics technologies~\cite{zhong18,wang19,wang19-b}.

\section*{Acknowledgement}

We thank P. Blasiak and M. Yang for useful discussion. This work is supported by National Research Foundation of Korea (2019M3E4A1079526, 2019R1A2C2006381) and a KIST research program (2E29580).


\end{document}